\documentstyle{pde96}
\Year{1996}
\newenvironment{remarks}{\begin{defi}{Remarks}}{\end{defi}}
\newenvironment{assumption}{\begin{defi}{Assumptions}}{\end{defi}}
\def\d{{\rm d}}
\def\e{{\rm e}}

\def\ass#1{\mbox{\rm\ref{ass}(#1)}}
\begin{document}
\Title{Rigorous results on Schr\"odinger operators with certain
Gaussian random potentials in multi-dimensional continuous space%
\footnote{Updated version of a contribution to the conference proceedings
{\em Differential equations, asymptotic analysis, and mathematical physics},
Eds.: M.\ Demuth and B.-W.\ Schulze, Akademie Verlag, Berlin (1997)
pp.\ 105--112}}
\Shorttitle{Gaussian random potentials}   
\By{{\sc Werner Fischer}, {\sc Thomas Hupfer}, {\sc Hajo Leschke},
{\sc Peter M\"uller}}
\Names{Fischer, Hupfer, Leschke, M\"uller} 
\Email{leschke@theorie1.physik.uni-erlangen.de}
\maketitle
%
%

\begin{abstract}
Schr\"odinger operators with certain Gaussian random potentials in 
multi-dimensional Euclidean space possess almost surely an absolutely 
continuous integrated density of states and no absolutely continuous 
spectrum at sufficiently low energies.
\end{abstract}

\newsection{Introduction} 

Schr\"odinger operators with Gaussian random potentials in 
$ d $-dimensional Euclidean space $ \R^d $, $ d\ge 1$, find 
wide-spread applications in physics. They are used, for example, 
to model aspects of disordered electronic systems such as 
heavily doped and highly compensated semiconductors \cite{BoEn84,ShEf84}.
Over several decades theoretical physicists have developed a good insight 
into the spectral characteristics of these operators
by combining intuitive ideas with approximation techniques and numerical 
studies. 
On the other hand there are still only few rigorous results
available \cite{Kir89,CaLa90,PaFi92}.
Our goal here is to present two new ones, which in the physics 
literature are often taken for granted.
More precisely, for $ \R^d $-homogeneous Gaussian random 
potentials with certain covariance functions
we are able to prove (i) the existence of the density of states, 
that is, the absolute continuity of the integrated density of states 
and (ii) the almost-sure absence of the absolutely continuous spectrum at
sufficiently low energies.

\newsection{Results}

Let $ (\Omega, {\cal A}, P) $ be a complete probability space and let
$ V:\, \Omega \times \R^{d} \rightarrow \R $, 
$ (\omega ,x)\mapsto V^{(\omega)}(x) $ be a real-valued homogeneous Gaussian 
random field indexed by $ \R^{d} $ with zero mean, 
$ \int_{\Omega } \!\d P(\omega )\, V^{(\omega)}(0) = 0 $, 
and covariance function
$ C(x):= \int_{\Omega }\! \d P(\omega )\, V^{(\omega)}(x)V^{(\omega)}(0) $ 
obeying $ C(0) > 0 $ and the

\begin{assumption}   \label{ass}
The covariance function $ C $ admits the representation
\begin{equation} \label{cerep}
C(x) = \int_{\R^d}\!\d^d y\; u(x+y) u(y)\,.
\end{equation}
Here $ \d^d x $ denotes the Lebesgue measure on $ \R^d $ and $ u $
is some real-valued Borel-measurable function on $ \R^d $ having the 
properties
\begin{itemize}\vspace{-.5ex} 
\item[(i)] Non-negativity: $ u(x) \ge 0 $ ~for all $ x \in\R^d $.
\item[(ii)] H\"older continuity: there exist constants $ a > 0 $ and 
$ \alpha \in ]0,1] $ such that
$$ | u(x+y) - u(x)| \le a |y|_{\infty }^\alpha $$ 
for all $ x, y \in \R^d $, where $ |x|_{\infty } := \sup\{ |x_{k}|:
\, 1\le k\le d\} $ is the usual supremum norm
of $ x =: (x_{1},\ldots , x_{d}) \in \R^{d} $.
\item[(iii)] Sufficiently fast decay at infinity: 
there exist constants $ b > 0 $ and 
$ \beta > (25d/2) -1 $ such that
$$ | u(x ) | \le \frac{b}{|x|_{\infty }^{\beta}} $$ 
for all $ x \in\R^d $.
\end{itemize}
\end{assumption}

\begin{remarks} \label{zuass}
\begin{itemize}\vspace{-.5ex} 
\item[(i)] The existence of the representation (\ref{cerep}) with 
some square-integrable function $ u $ is equivalent to the fact that 
$ C $ is the Fourier transform of a finite measure which is absolutely 
continuous with respect to $ \d^d x $. 
\item[(ii)] Assumptions \ref{ass} imply that 
$ C $ is a non-negative H\"older continuous function on $ \R^d $
which tends sufficiently fast to zero at infinity. Due to the H\"older 
continuity of $ C $ it follows 
from \cite[Thm.\ 3.2.2]{Fer75} that there is a version of $ V $
which is jointly measurable with respect to the
sigma-algebra $ {\cal A} $ and the Borel sigma-algebra of $ \R^d $. 
Moreover, the realisations of this version are $ P $-almost surely 
continuous \cite[Thm.\ 4.1.1]{Fer75}. The vanishing of $ C $ at infinity
implies mixing and hence ergodicity of $ V $.
Taken together, a compromise is required between local 
dependence and global independence of $ V $.
\item[(iii)] 
Since $ C(0) > 0 $ and $ C $ is continuous there is a length $ \ell > 0 $
such that 
\begin{equation} \label{gamma}
\gamma := \inf_{|x|_{\infty }< \ell/2} \{ C(x)/C(0) \} > 0 \,.
\end{equation}
\item[(iv)]
An example of a covariance function satisfying Assumptions \ref{ass}
is the Gaussian
\begin{equation} \label{gauss}
C(x) = \sigma ^2 \exp\{ -x^2/2\xi^2\}\,, \qquad \sigma ,\xi > 0 \,.
\end{equation}
\end{itemize}
\end{remarks}

\vspace{2.5ex}
With the Gaussian random field $ V $ and a bounded open (hyper-) cube 
$ \Lambda \subset \R^d $ we associate the finite-volume random 
Schr\"odinger operator 
\begin{equation} \label{finham}
H_{\Lambda ,{\rm X}}^{(\omega )}      
:= \left( - {\textstyle\frac{1}{2}} \,\Delta  + V^{(\omega )}
\right)_{\Lambda ,{\rm X}} \; , \qquad {\rm X}={\rm D} {\rm ~or~} 
{\rm X}={\rm N} 
\end{equation}
acting on the Hilbert space $ {\rm L}^{2} (\Lambda ) $ of complex-valued 
functions on $ \Lambda $ which are square-integrable over $ \Lambda $
with respect to $ \d^{d}x $.
According to the subscript $ {\rm X} $ the functions in the domain of 
$ H_{\Lambda ,{\rm X}}^{(\omega )} $ obey either a Dirichlet or a Neumann 
condition on the boundary of $ \Lambda $.
Here, as usual, $ \Delta := \sum_{k=1}^{d} \partial^{2}/\partial x_{k}^{2} $ 
is the $ d $-dimensional Laplacian and $ V $ appears as a random 
potential which acts as a multiplication operator.
We also consider the infinite-volume random Schr\"odinger operator 
\begin{equation} \label{infinham}
H^{(\omega )} :=  - {\textstyle\frac{1}{2}} \,\Delta  + V^{(\omega )}
\end{equation}
which acts on the Hilbert space $ {\rm L}^{2} (\R^d ) $.

\begin{remarks}
According to standard arguments it is known \cite{Kir89,CaLa90,PaFi92}
that
\begin{itemize}\vspace{-.5ex}
\item[(i)] the finite-volume operator $ H_{\Lambda ,{\rm X}}^{(\omega )} $
is self-adjoint on the domain of the free operator
$ - \frac{1}{2}\,\Delta _{\Lambda ,{\rm X}} $ for $P$-almost all $ \omega $.
The infinite-volume operator $ H^{(\omega )} $
is essentially self-adjoint for $P$-almost all $ \omega $
on the dense subspace 
$ {\mathcal{C}}_{0}^{\infty}(\R^{d}) \subset {\rm L}^{2}(\R^{d}) $, consisting 
of arbitrarily often differentiable 
complex-valued functions with compact support in $ \R^{d} $.
\item[(ii)]
the spectrum of $ H_{\Lambda,{\rm X}}^{(\omega )} $ is 
discrete for $ P $-almost all $\omega $. Hence the finite-volume
integrated density 
of states $ N_{\Lambda,{\rm X} }^{(\omega )}(E) $, defined as the number of 
eigenvalues of $ H_{\Lambda,{\rm X}}^{(\omega )} $ 
which are smaller than $ E $ and counted with their multiplicities,
exists for $ P $-almost all $\omega $.
\item[(iii)] there is a non-random unbounded distribution function $ N $ 
on $ \R $, called the infinite-volume integrated density of states,
and a set $ \Omega _{0}\in {\cal A} $ of maximal probability,
$ P(\Omega _{0}) =1 $, such that the macroscopic-limit relation
\begin{equation} \label{Nmak}
N(E) = \lim_{\Lambda  \uparrow \R^d} 
\frac{N_{\Lambda, {\rm X}}^{(\omega )}(E)}{|\Lambda |} 
\end{equation}
holds for both boundary conditions $ {\rm X} $, 
for all $ \omega\in\Omega _{0} $ and for all those
$ E\in \R $ which are continuity points of $ N $. 
Here $ |\Lambda | $ denotes the volume of the cube $ \Lambda \subset 
\R^d $ measured by $ \d^d x $. 
The infinite-volume integrated density of states may also be expressed 
as an average of the localised spectral projection of the infinite-volume 
operator according to 
\begin{equation} \label{spurformel}
N(E) = \frac{1}{|\Lambda |} \int_{\Omega }\!\d P(\omega )\;
{\rm Tr}\left(\chi _{\Lambda }^{\phantom{f}} \Theta (E -H^{(\omega )})
\chi _{\Lambda }^{\phantom{f}}\right)\,.
\end{equation}
Here $ \Lambda\subset\R^d  $ is an arbitrary bounded cube with 
$ |\Lambda | > 0 $. Its indicator function 
$ \chi _{\Lambda }^{\phantom{f}} $ appears in (\ref{spurformel})  
as a multiplication operator.
Furthermore, $ \Theta $ denotes Heaviside's unit-step function and 
$ {\rm Tr} $ stands for the trace over $ {\rm L}^2(\R^d) $.
The leading asymptotic low- and high-energy behaviour of $ N $ is given by
\begin{eqnarray}
\lim_{E\to -\infty }\frac{\ln N(E)}{E^2} &=& -\frac{1}{2C(0)}\,,
\label{lowlimit} \\
\lim_{E\to \infty }\frac{N(E)}{E^{d/2}} &=& 
\frac{1}{(2\pi )^{d/2}\,\Gamma (1+ d/2)}\,, \label{highlimit}
\end{eqnarray}
where $ \Gamma $ denotes Euler's gamma function.
\item[(iv)] 
the topological support of the measure associated with $ N $ 
equals the real line $ \R $ and 
coincides with the spectrum $ \sigma (H^{(\omega )}) $ of the infinite-volume  
operator for $P$-almost all $ \omega $.
The spectral components, the absolutely continuous, the singular 
continuous and the pure point spectrum, arising in the Lebesgue 
decomposition $ \sigma (H^{(\omega )}) = \sigma_{\rm ac} (H^{(\omega )})\cup 
\sigma_{\rm sc} (H^{(\omega )})\cup \sigma_{\rm pp} (H^{(\omega )}) $ are 
also non-random closed sets for $P$-almost all $ \omega $.
\end{itemize}
\end{remarks}

\vspace{.2cm}
As our first result we state a Wegner estimate for 
the finite-volume situation in

\begin{theorem} \label{wegner}
Let the finite-volume operator $ H_{\Lambda ,{\rm X}}^{(\omega )} $ 
be defined as in 
{\rm (\ref{finham})} with a homogeneous Gaussian random potential 
satisfying Assumptions {\rm\ref{ass}}. 
Then for every energy $ E \in \R $ there exists a 
constant $ 0< W(E) <\infty $, which is independent of $  \Lambda $ and 
$ {\rm X} $, 
such that the finite-volume integrated density of states 
$ N_{\Lambda, {\rm X} }^{(\omega )} $ obeys    
\begin{equation} \label{wegnerglg}
\int_{\Omega }\!\d P(\omega ) \left| 
N_{\Lambda, {\rm X} }^{(\omega )}(E_{1}) -
N_{\Lambda, {\rm X} }^{(\omega )}(E_{2}) \right| \le |\Lambda |\, 
|E_{1}-E_{2}| \, W(E)
\end{equation}
for all $ E_{1},E_{2} \le E $ and all bounded open 
cubes $ \Lambda \subset \R^d $ with $ |\Lambda | \ge \ell^d  $.
(The length $ \ell $ is defined in Remark \mbox{\rm\ref{zuass}(iii)}.)
\end{theorem}

\vspace{.5cm}
The Lipschitz continuity (\ref{wegnerglg}) of the averaged finite-volume 
integrated density of states implies by
the non-randomness of the infinite-volume integrated density 
of states and Fatou's lemma the following

\begin{corollary}    \label{corzuwegner}
Under the assumptions of Theorem {\rm \ref{wegner}} 
the infinite-volume integrated density of states $ N $ is
absolutely continuous on any bounded interval and its derivative, 
the density of states, is locally bounded in the sense that
\begin{equation}
0\le \frac{\d N(E)}{\d E} \le W(E)
\end{equation}
for Lebesgue-almost all $ E\in\R $.
\end{corollary}

\begin{remarks}
\begin{itemize}\vspace{-.5ex}
\item[(i)] 
Theorem \ref{wegner} and Corollary \ref{corzuwegner} are proved in
\cite{FiHu96} under weaker assumptions than the ones used here.
Actually it is shown there that Theorem \ref{wegner} is a 
consequence of a Wegner estimate which holds for 
all continuum Schr\"odinger operators whose random potential admits a certain 
one-parameter decomposition. For the present case the basic idea is to 
write 
\begin{equation} \label{decomp}
V^{(\omega )}(x) =: U^{(\omega )}(x) + V^{(\omega )}(0) \;
\frac{C(x)}{C(0)} \,.
\end{equation}
Since the Gaussian random variable $ V(0) $ is stochastically 
independent of the (non-homogeneous) Gaussian random field $ U $,
spectral averaging with respect to $ V(0) $ is easily performed and 
one may adapt the line of reasoning laid down in \cite{CoHi94a} 
to prove Theorem \ref{wegner}. 
\item[(ii)] It follows from the proof of Theorem \ref{wegner} 
that the Wegner constant $ W(E) $ may be taken as
\begin{equation} \label{wegnerkonst}
W(E) = \frac{\exp\{t E + t ^2 C_{E}/2\}}%
{\sqrt{2\pi C(0)}\; b_{E}}\,\Big( 2 \ell_{E}^{-1} + 
(2\pi t )^{-1/2}\Big)^d\,,
\end{equation}
where $ t > 0 $ is arbitrary and may be considered as a variational 
parameter. In (\ref{wegnerkonst}) we are using the constants
\begin{eqnarray}
\ell_{E} &:=& \inf\{ |E|^{-1/2}, \ell \}\,, \\
b_{E} &:=&  \inf_{|x|_{\infty } < \ell_{E}/2}
\{C(x)/C(0)\} \ge  \gamma \,,\\
C_{E} &:=& C(0) (2 - b_{E}^2)\,.
\end{eqnarray}
The choice $ t = (2 C_{E})^{-1} (-E + \sqrt{E^2 + 2C_{E}/\pi })  $ shows 
that $ W $ has the same low- and high-energy behaviour as $ N $ 
except that the constant on the right-hand side in (\ref{highlimit})
is to be replaced by $ 3^d\,\e^{1/2\pi } /\sqrt{2\pi C(0)} $.
\end{itemize}
\end{remarks}

\vspace{.2cm}
As our second result we state for the infinite-volume situation
the almost-sure absence of the absolutely continuous 
spectrum at sufficiently low energies in

\begin{theorem} \label{lokalisierung}
Let the infinite-volume operator $ H^{(\omega )} $ be defined as in 
{\rm (\ref{infinham})} with a homogeneous Gaussian random potential 
satisfying Assumptions {\rm\ref{ass}}. Then there exists a finite energy 
$ E_{0} <0 $ such that
\begin{equation}
\sigma_{\rm ac} (H^{(\omega )})\; \cap\; ]-\infty ,\,E_{0}] = \emptyset
\qquad {\rm\mbox{for $ P$-almost all $\omega $.}}
\end{equation}
\end{theorem}

\begin{remarks}
\begin{itemize}  \vspace{-.5ex}
\item[(i)] The heart of the proof of Theorem \ref{lokalisierung}
is a multi-scale analysis in the spirit of the
fundamental work \cite{FrSp83}. 
Its technical realisation is patterned after 
\cite{DrKl91} and \cite{CoHi94a} in order to cope with a correlated 
random potential and a continuous space, respectively. 
A key r\^ole in the proof is played by the Wegner estimate
Theorem \ref{wegner} above. 
In case of a Gaussian random field with a sufficiently fast 
decaying strong-mixing coefficient
a proof of the statement of Theorem \ref{lokalisierung} is outlined in 
\cite{FiLe96a} and will be completed in \cite{FiLe96b}.
In case of the present assumptions on $ V $ a more refined strategy is 
required in that different Gaussian random fields $ V_{L_{j}} $ are used 
on the different length 
scales $ L_{j} $ of the multi-scale analysis. The correlation radius of 
$ V_{L_{j}} $ is of order $ L_{j-1} $ and in the macroscopic limit 
$ V_{L_{j}} $ tends sufficiently fast to the given random field $ V $ 
in $ P $-mean square. 
This is why Assumption \ass{iii} requires such a strong decay of $ u $.
Again we defer the details of the proof to \cite{FiLe96b}.
\item[(ii)] So far we have not yet been able to rule out the 
singular continuous spectrum for sufficiently low energies. But we have
some hope that the decomposition (\ref{decomp}) of $ V $ may be used to 
proceed along the lines in \cite{SiWo86,How87,CoHi94a}.
\item[(iii)] Theorem \ref{lokalisierung} applies in particular to a
Gaussian random field with the Gaussian covariance function (\ref{gauss}). 
For one space dimension this 
is even an example of a so-called deterministic stochastic process
\cite[ch.\ IV, \S 9, Thm.\ 7]{IbRo78}.
\item[(iv)]  The absence of the (absolutely) continuous spectrum is in 
general not sufficient to imply vanishing transport coefficients. Confer
the recent critical discussions \cite{Sim90,RiJi95,BaCo96,Com96}
on the relation between quantum dynamics and spectral characteristics.
\end{itemize}
\end{remarks}

\begin{acknowledgements}
We are grateful to P.\ D.\ Hislop for stimulating discussions.
One of us (H.L.) wishes to thank M.\ Demuth and B.-W.\ Schulze
for inviting him to Potsdam and for the inspiring atmosphere 
created there.
This work was partially supported by the Deutsche 
Forschungsgemeinschaft and by the Human Capital and Mobility
Programme ``Polarons, bi--polarons and excitons. Properties and 
occurrence in new materials.'' of the European Community.
\end{acknowledgements}


\noindent{\bf Addresses:}

{\sc Werner Fischer; Thomas Hupfer; Hajo Leschke; Peter M\"uller,} 
Universit\"at Erlangen-N\"urnberg, Institut f\"ur Theoretische Physik,
Staudt\-stra{\ss}e~7, D-91058 Erlangen, Germany.


\begin{references}{CCCM}
%
\bibitem[BCM]{BaCo96}
Barbaroux, J.-M., Combes, J.-M. and Montcho, R.:
Remarks on the relation between quantum dynamics and fractal spectra.
Preprint mp-arc 96-293, 1996
%
\bibitem[BEE+]{BoEn84}
Bonch-Bruevich, V. L., Enderlein, R., Esser, B., 
Keiper, R., Mironov, A. G. and Zvyagin, I. P.:
Elektronentheorie ungeordneter Halbleiter.
VEB Deutscher Verlag der Wissenschaften, Berlin, 1984 
%
\bibitem[CL]{CaLa90}
Carmona, R. and Lacroix, J.:
Spectral theory of random Schr\"odinger operators.
Birkh\"auser, Boston, 1990
%
\bibitem[C]{Com96}
Combes, J.-M.:
Dynamics of wave-packets associated to unusual spectras: 
a review of recent results.
In: Demuth, M. and Schulze, B.-W. (ed.):
Differential equations, asymptotic analysis, and mathematical physics.
Akademie Verlag, Berlin, 1997, pp.54--62
%
\bibitem[CH]{CoHi94a}
Combes, J.-M. and Hislop, P. D.:
Localization for some continuous, random Hamiltonians in $ d $-dimensions.
J. Funct. Anal. {\bf 124} (1994), 149--180
%
\bibitem[DK]{DrKl91} 
von Dreifus, H. and Klein, A.:
Localisation for random Schr\"odinger operators with correlated 
potentials.
Commun. Math. Phys. {\bf 140} (1991), 133--147
%
\bibitem[F]{Fer75}
Fernique, X. M.:
Regularit\'e des trajectoires des fonctions al\'eatoires Gaussiennes.
In: Hennequin, P.-L. (ed.): 
Ecole d'Et\'e de Probabilit\'es de Saint-Flour IV - 1974. 
Lecture Notes in Mathematics vol. 480, 
Springer, Berlin, 1975, pp. 1--96
%
\bibitem[FHLM]{FiHu96}
Fischer, W., Hupfer, T., Leschke, H. and M\"uller, P.:
Existence of the density of states for multi-dimensional
continuum Schr\"odinger operators with Gaussian random potentials.
Preprint {mp-arc 97-245}. To appear in Commun. Math. Phys. (1997)
%
\bibitem[FLM1]{FiLe96a}
Fischer, W., Leschke, H. and M\"uller, P.:
Towards localisation by Gaussian random potentials in multi-dimensional
continuous space.
Lett. Math. Phys. {\bf 38} (1996), 343--348
%
\bibitem[FLM2]{FiLe96b}
Fischer, W., Leschke, H. and M\"uller, P.:
in preparation, to be submitted to J. Stat. Phys.
%
\bibitem[FS]{FrSp83}
Fr\"ohlich, J. and Spencer, T.:
Absence of diffusion in the Anderson tight binding model for large 
disorder or low energy.
Commun. Math. Phys. {\bf 88} (1983), 151--184 
%
\bibitem[H]{How87}
Howland, J. S.:
Perturbation theory of dense point spectra.
J. Funct. Anal. {\bf 74} (1987), 52--80 
%
\bibitem[IR]{IbRo78} 
Ibragimov, I. A. and Rozanov, Y. A.:
Gaussian random processes. 
Springer, New York, 1978
%
\bibitem[K]{Kir89}
Kirsch, W.:
Random Schr\"odinger operators, a course. In: Holden, H. and Jensen, A. 
(eds.): Schr\"odinger operators. Lecture Notes in Physics vol. 345,
Springer, Berlin, 1989, pp. 264--370
%
\bibitem[PF]{PaFi92}
Pastur, L. and Figotin, A.:
Spectra of random and almost-periodic operators.
Springer, Berlin, 1992
%
\bibitem[RJLS]{RiJi95}
del Rio, R., Jitomirskaya, S., Last, Y. and Simon, B.:
What is localization?
Phys. Rev. Lett. {\bf 75} (1995), 117--119 
%
\bibitem[SE]{ShEf84}
Shklovskii, B. I. and Efros, A. L.:
Electronic properties of doped semiconductors.
Springer, Berlin, 1984
%
\bibitem[S]{Sim90}
Simon, B.:
Absence of ballistic motion.
Commun. Math. Phys. {\bf 134} (1990), 209--212 
%
\bibitem[SW]{SiWo86}
Simon, B. and Wolff, T.:
Singular continuous spectrum under rank one perturbations
and localization for random Hamiltonians.
Commun. Pure Appl. Math. {\bf 39} (1986), 75--90
%
\end{references}
\end{document}